\documentclass[reprint, amsmath, amssymb, aip, apl]{revtex4-1}
\usepackage{graphicx}
\usepackage{amsmath}
\usepackage{amssymb}
\usepackage{bm}
\usepackage{physics}
\usepackage{comment}
\usepackage{textcomp}
\usepackage{times}
\usepackage{color}
\usepackage{soul}
\usepackage{siunitx}
\usepackage{titlesec}

\titleformat{\part}
  {\normalfont\fontsize{12}{15}\bfseries}{\thepart}{1em}{}

\begin{document}

\title{Nitrogen isotope effects on boron vacancy quantum sensors in hexagonal boron nitride}

\author{Kento Sasaki}
\email{kento.sasaki@phys.s.u-tokyo.ac.jp}
\affiliation{Department of Physics, The University of Tokyo, 7-3-1 Hongo, Bunkyo-ku, Tokyo 113-0033, Japan}
\author{Takashi Taniguchi}
\affiliation{Research Center for Materials Nanoarchitectonics, National Institute for Materials Science, 1-1 Namiki, Tsukuba 305-0044, Japan}
\author{Kensuke Kobayashi}
\email{kensuke@phys.s.u-tokyo.ac.jp}
\affiliation{Department of Physics, The University of Tokyo, 7-3-1 Hongo, Bunkyo-ku, Tokyo 113-0033, Japan}
\affiliation{Institute for Physics of Intelligence, The University of Tokyo, 7-3-1 Hongo, Bunkyo-ku, Tokyo 113-0033, Japan}
\affiliation{Trans-scale Quantum Science Institute, The University of Tokyo, 7-3-1 Hongo, Bunkyo-ku, Tokyo 113-0033, Japan}

\date{\today}

\begin{abstract}
There has been growing interest in studying hexagonal boron nitride (hBN) for quantum technologies. 
Here, we investigate nitrogen isotope effects on boron vacancy (V$_\text{B}$) defects, one of the candidates for quantum sensors, in $^{15}$N isotopically enriched hBN synthesized using a metathesis reaction. 
The Raman shifts are scaled with the reduced mass, consistent with previous work on boron isotope enrichment.
We obtain nitrogen isotopic composition-dependent magnetic resonance spectra of V$_\text{B}$ defects and determine the magnitude of the hyperfine interaction parameter of $^{15}$N spin to be $64~\mathrm{MHz}$. 
Our investigation provides a design policy for hBNs for quantum sensing.
\end{abstract}

\maketitle

Localized electron spins in solids, such as those in color centers or quantum dots, are the promising platform of quantum technologies.
In most cases, they couple with surrounding nuclear spins; thus, controlling the nuclear spins and their influence is essential.
The isotope enrichment technique has great potential to address this issue\cite{Itoh2014}.
For example, the electron spin coherence time can be improved by enriching nuclear-spin-free isotopes\cite{Balasubramanian2009,Ishikawa2012,Ohashi2013,Muhonen2014,Veldhorst2014,Kleinsasser2016}, or the electron spin qubit can be labeled by isotopes with low natural composition ratios\cite{Rabeau2006,vanDam2019}.
In designing such an isotopically purified platform, it is crucial not only to synthesize isotopically controlled materials but also to estimate the isotopic composition and determine the hyperfine interaction (HFI) parameters of nuclear spins of the isotopes\cite{Rabeau2006,vanDam2019}.

\begin{figure}
\begin{center}
\includegraphics{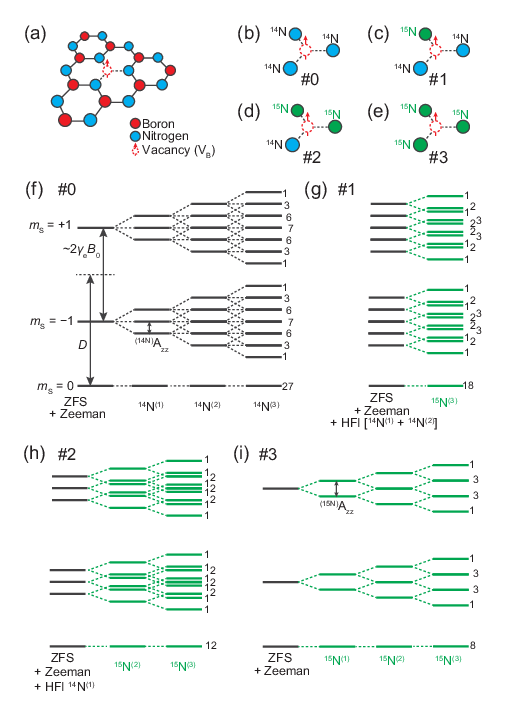}
\caption{
Structure and energy levels of boron vacancy (V$_\text{B}$) defects.
(a) Structure of V$_\text{B}$ defects.
We distinguish the V$_\text{B}$ defect by the number of $^{15}$N among the nearest nitrogen atoms: (b)\#0, (c)\#1, (d)\#2, (e)\#3.
The energy level splitting of \#0, \#1, \#2, and \#3 are shown in (f), (g), (h), and (i), respectively.
The number at the right of each level indicates the number of the degeneracy.
Minor energy shifts such as nuclear spin's Zeeman effect and quadrupole interaction are ignored [see Eq.~(\ref{ham})].
ZFS: zero-field splitting, HFI: hyperfine interaction.
\label{fig1}
}\end{center}
\end{figure}

Recently, it has been discovered that electron spins of boron vacancy (V$_\text{B}$) defects in hexagonal boron nitride (hBN) can be used as quantum sensors even at room temperature\cite{Gottscholl2020,Gottscholl2021,Huang2022,Healey2022,Kumar2022,Sasaki2023}.
A V$_\text{B}$ defect has a structure in which a boron atom in hBN is replaced by a vacancy [Fig.~\ref{fig1}(a)].
Its electron spin is localized around the vacancy site and is significantly affected by the three nearest nitrogen nuclear spins.
Stable isotopes of nitrogen are $^{14}$N and $^{15}$N.
The natural composition ratio of $^{14}$N is 99.6\%, and $^{15}$N is almost nonexistent (0.4\%).
The nuclear spin is one of the major differences between these isotopes.
Since $^{15}$N spin ($I=1/2$) is only half of the $^{14}$N spin ($I=1$), V$_\text{B}$ defects in $^{15}$N isotopically enriched hBN have fewer energy levels than in non-treated hBN.
The fewer levels and the higher the occupancy, the stronger and less overlapping each resonance signal will be.
They increase the sharpness of the resonance signal and thus lead to higher sensitivity.
However, there are few reports on the isotope enrichment of hBN, most of which are related to boron isotopes\cite{Vuong2017,Cusc2018,Haykal2022,Janzen2023}.

%Here we
Here, we investigate nitrogen isotope-enriched hBN and observe nitrogen isotope effects on the optically detected magnetic resonance (ODMR) spectrum of V$_\text{B}$ defects.
We synthesized the isotopically controlled hBN crystals using a metathesis reaction under high pressure\cite{Chen2020,TaniguchiXXXX} with commercially available $^{15}$NH$_4$Cl.
The Raman shifts of the samples are scaled with their reduced mass, which is the effective mass for an equivalent one-body problem of the two-body vibration problem for boron and nitrogen atoms, consistent with previous work on boron isotope enrichment.
We perform ODMR measurements of V$_\text{B}$ defects created by helium ion implantation and determine the magnitude of the HFI parameter of $^{15}$N spin to be $64~\mathrm{MHz}$.
The observed significant modification of resonance spectra due to $^{15}$N isotope enrichment will help improve sensitivity, control fidelity, and precise positioning of quantum sensors.
Our investigation provides guidance for the material design of hBNs for quantum technologies.

First, we describe the influence of nitrogen spins on an electron spin ($S = 1$) of a V$_\text{B}$ defect.
In magnetic field sensing, an external magnetic field of several mT in the direction of the symmetry axis ($z$) of the V$_\text{B}$ defect is often applied~\cite{Huang2022,Healey2022,Kumar2022,Sasaki2023} to mitigate the sensitivity suppression due to the strain.
In that condition, the spin Hamiltonian can be approximated as~\cite{Gao2022},
\begin{align}
\hat{H} &\sim D \hat{S}_z^2 + \gamma_e \bm{B}_z \cdot \hat{S}_z + \sum_{j=1}^{3} A_{\text{zz},(j)} \hat{S}_z \hat{\bm{I}}_{z,(j)},
\label{ham}
\end{align}
where $\hat{S}_{z}$ is the electron spin ($S=1$) operator in the $z$ direction, $D$ is the zero field splitting, $\gamma_e = 28~\mathrm{MHz/mT}$ is the gyromagnetic ratio of the electron spin, $B_z$ is the magnetic field strength, $j (= 1,2,3)$ is a label of nearest-neighbor nitrogen site, $A_{zz,(j)}$ is the HFI parameter, and $\hat{I}_{z,(j)}$ is the nuclear spin operator in the $z$ direction [see Supplemental Information (SI)].
Here, we ignore the nuclear spin's Zeeman effect and the quadrupole moment\cite{Gracheva2023}, which are much smaller than the HFI parameter in the case of the $^{14}$N spin.
In this study, we determine the $|A_{zz}|$ of $^{15}$N spin, $|^\text{(15N)}A_{zz}|$, that has vital contributions in this quantum sensing condition.

Next, we show a model of the expected ODMR spectrum.
When Eq.~(\ref{ham}) is valid, electron and nuclear spins are quantized in the $z$ direction.
The resonance frequency corresponding to the electron spin transition $m_S = 0 \leftrightarrow \pm1$ can be expressed as
\begin{align}
f_{\pm1}(m_{I,(1)},m_{I,(2)},m_{I,(3)}) &\sim f_{\pm1,0} \pm \sum_{j=1}^{3} A_{\text{zz},(j)} m_{I,(j)},
\label{eqf}
\end{align}
where $f_{\pm1,0} = D \pm \gamma_e B_z$ is the resonance frequency in the absence of nuclear spins and $m_{I,(j)}$ is the magnetic quantum number of nuclear spins at site $j$ which can take the values $m_I=-1,0,+1$ for $^{14}$N spin ($m_I=-1/2,+1/2$ for $^{15}$N spin).
Assuming that the nuclear spins are unpolarized and each resonance signal has the same amplitude and line width, the ODMR spectrum is given by
\begin{align}
R = 1 - \frac{C}{N_\text{level}} \sum L( f_{\pm1}(m_{I,(1)},m_{I,(2)},m_{I,(3)}), d\nu),
\label{odmreach}
\end{align}
where $C$ is the signal amplitude and $L(f,d\nu)$ is the Lorentzian with a center frequency $f$ and a full width at half maximum $d\nu$.
$N_\text{level}$ is the number of possible nuclear states of the nearest-neighbor nitrogen spins $(m_{I,(1)},m_{I,(2)},m_{I,(3)})$, and the summation symbol means summing concerning those states, which will be explained in detail below.

The resonance spectrum of a V$_\text{B}$ defect [Eq.~(\ref{odmreach})] depends on the number $n$ of $^{15}$N among the nearest nitrogen atoms.
We distinguish V$_\text{B}$ defects by \#$n$, as shown in Figs.~\ref{fig1}(b--e).
The energy level splittings of these defects are shown in Figs.~\ref{fig1}(f--i).
Since $^{14}$N spins can take three states ($m_I=-1,0,+1$), whereas $^{15}$N spins can take only two states ($m_I=-1/2,+1/2$), $N_\text{level}$ of \#0, \#1, \#2 and \#3 are 27($=3^3$), 18($=3^2\times2$), 12($=3\times2^2 $), and 8($=2^3$), respectively.
To the extent that Eq.~(\ref{ham}) is satisfied, all states belonging to $m_S=0$ and some of the states belonging to $m_S=\pm1$ are degenerated.
In the case of $m_S=-1$ of \#0 (\#3), there are 7 (4) states whose energies are distinguished by the total nuclear spin quantum number, $m_{I,tot} = \sum_{j=1}^{3} m_{I,(j)}$.
Specifically, the degeneracy of energy states $m_{I,tot}$ = -3, -2, -1, 0, +1, +2, and +3 (-3/2, -1/2, +1/2, and 3/2) are 1, 3, 6, 7, 6, 3, and 1 (1, 3, 3, and 1), respectively [see Figs.~\ref{fig1}(f) and (i)].
The occupancy of the state with the largest degeneracy is 26\% ($=7/27$) for \#0 and 38\% ($=3/8$) for \#3.

The distances between energy states (= resonance lines) depend on the magnitude of the HFI parameter $|A_{zz}|$ of $^{14}$N and $^{15}$N spins [Eq.~(\ref{eqf})].
The gyromagnetic ratio, the magnetic moment per unit spin angular momentum, is $\gamma_\text{14N} = 3.077~\mathrm{kHz/mT}$ for $^{14}$N spin and $\gamma_\text{15N} = -4.316~\mathrm{kHz/mT}$ for $^{15}$N spin.
Since the HFI parameter is proportional to the magnetic moment, the spectral separation for $^{15}$N isotope-enriched hBN is expected to be 1.4 times larger than the conventional case.
In the conventional case, the hyperfine interaction and line width limited by boron nuclear spins are comparable~\cite{Haykal2022}, resulting in a significant spectrum overlap.
It degrades the sharpness/slope of the ODMR spectrum and makes it challenging to manipulate electron spins selectively to desired nuclear spin states~\cite{Gu2023}.
It is also unfavorable for the magnetic field sensitivity, which is proportional to the slope of the ODMR spectrum (see SI).
In the $^{15}$N case, the increased occupancy and enhanced distance between resonance lines will help to reduce the overlap and sharpen ODMR spectra; thus, they are advantageous to improve magnetic field sensitivity and control fidelity.
In this work, we will demonstrate the nitrogen isotope effects described above, such as a reduced number of resonance lines and enhanced separation.

When measuring an ensemble of V$_\text{B}$ defects, the signals of \#0 to \#3 are averaged.
Specifically, the expected ODMR spectrum is given by,
\begin{align}
R_\text{tot} &= P_0 R_0 + P_1 R_1 + P_2 R_2 + P_3 R_3,
\label{odmrtot}
\end{align}
where $R_n$ is the ODMR spectrum $R$ of \#$n$ [Eq.~(\ref{odmreach})] and $P_n$ is the fraction of \#$n$ in all V$_\text{B}$ defects.
When $^{15}$N isotopic composition, $p_{15}$, is spatially uniform, then $P_0 = (1 - p_{15})^3, P_1 = 3(1 - p_{15})^2 p_{15}, P_2 = 3(1 - p_{15 }) p_{15}^2$, and $P_3 = p_{15}^3$.

Here, we describe the preparation of $^{15}$N isotopically enriched hBN crystal.
We verify the metathesis reaction process under high pressure\cite{Chen2020,TaniguchiXXXX} using commercially available ammonium chloride $^{15}$NH$_4$Cl ($^{15}$N: 99~\%, Cambridge Isotope Laboratories) reagents as a raw material; NaBH$_4$ + $^{15}$NH$_4$Cl = B$^{15}$N + NaCl + 4H$_2$.
By continuing the above reaction for about 30~hours, we obtained hBN crystals, which are expected to be close to the perfect $^{15}$N isotopic composition (hB$^{15}$N).
Other hBN single crystals of about 1~mm are obtained using Ba-BN as a solvent system~\cite{Taniguchi2007}, where hBN sources are grown within the molten solvent through dissolution and precipitation.
In this case, the nitrogen isotope enrichment in the resulting crystals (hB$^{14+15}$N) is not 100\% because nitrogen in Ba-BN solvents has a natural isotopic composition.
The $^{15}$N isotopic composition of hB$^{14+15}$N is determined by secondary ion mass spectrometry (SIMS) as $59.5\pm0.3$~\%.
In addition, hBN crystal with a natural composition ratio (hB$^{14}$N) is used for comparison.
To simplify the analysis, we approximate $p_{15}$ of hB$^{14}$N, hB$^{14+15}$N, and hB$^{15}$N as $0~\%$, $60~\%$, and $100~\%$, respectively.

\begin{figure}
\begin{center}
\includegraphics{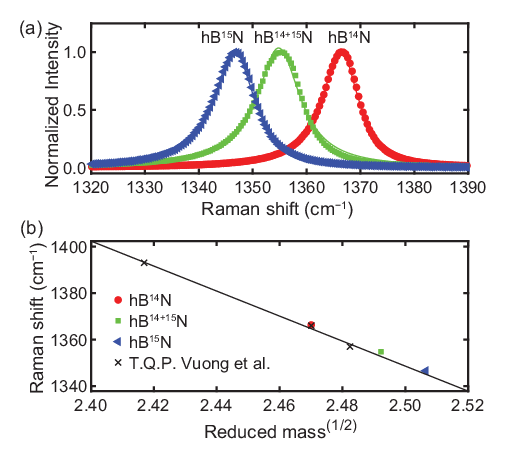}
\caption{
Isotope effects on the phonon energy.
(a) The Raman scattering spectrum of each sample.
The solid lines are the Lorentzian fit to obtain the Raman shift.
(b) Relationship between the square root of the reduced mass and the Raman shift.
The circle, square, and triangle are the results for hB$^{14}$N, hB$^{14+15}$N, and hB$^{15}$N, respectively.
The crosses are the result of previous work on boron isotopically enriched crystals~\cite{Vuong2017}, and the solid line is the linear fit of them.
\label{fig2}
}\end{center}
\end{figure}

From now on, we will describe the experimental results.
All the measurements in this work are performed at room temperature.
First, we investigate the isotope effect on the phonon energy due to changes in the reduced mass using a Raman microscope (Nanophoton RAMAN-FM-UTM).
In previous works on boron isotope enrichment~\cite{Cusc2018,Vuong2017}, it has been shown that the phonon energy scales with the square root of the reduced mass.
Figure~\ref{fig2}(a) shows the obtained Raman scattering spectra.
The sample with a natural composition ratio, hB$^{14}$N, has a Raman shift of $1366.3$~cm$^{-1}$.
This value is consistent with the previous work~\cite{Stenger2017}.
In contrast, the Raman shifts for hB$^{14+15}$N and hB$^{15}$N are $1354.8$~cm$^{-1}$ and $1346.6$~cm$^{-1}$, respectively.
Clearly, the Raman shift decreases with increasing $^{15}$N isotopic composition, i.e. increasing reduced mass.

To quantitatively evaluate this behavior, we show the relationship between Raman shift and reduced mass in Fig.~\ref{fig2}(b).
By analyzing the result of Ref.~\onlinecite{Vuong2017}, we obtain,
\begin{align}
\Delta \nu_\text{r} &\sim -537 \mu^{1/2} + 2691,
\label{shift}
\end{align}
where $\Delta \nu_\text{r}$ is the Raman shift (unit $\mathrm{cm^{-1}}$), and $\mu$ is the reduced mass (no unit).
The crosses and the solid line in Fig.~\ref{fig2}(b) are the results of Ref.~\onlinecite{Vuong2017} and Eq.~(\ref{shift}), respectively.
The deviation between them is as slight as about 1~cm$^{-1}$.
Since our results agree with Eq.~(\ref{shift}) within the error of about 2~cm$^{-1}$, we confirm that our nitrogen isotope enrichment is successful.

\begin{figure}
\begin{center}
\includegraphics{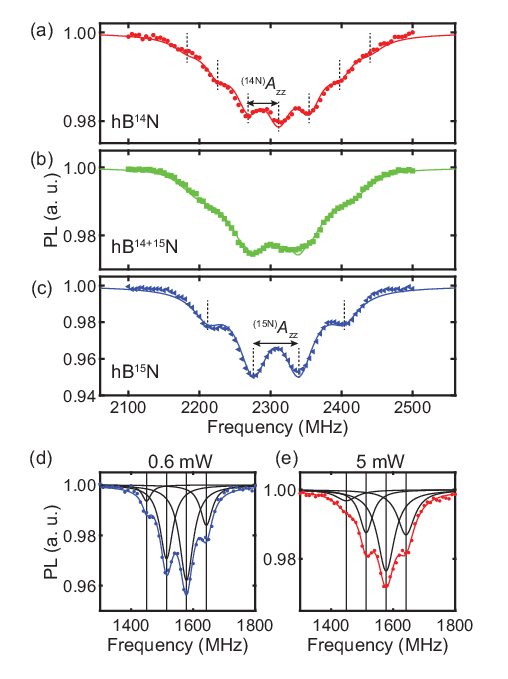}
\caption{
Nitrogen isotope effects on V$_\text{B}$ defects.
ODMR spectra of (a) hB$^{14}$N, (b) hB$^{14+15}$N, and (c) hB$^{15}$N at $B_z \sim 40~\mathrm{mT}$.
The vertical axis is the photoluminescence intensity normalized by that without microwave application.
The solid lines are the fitting results using Eq.~(\ref{odmrtot}). 
ODMR spectra of hB$^{15}$N at $B_z\sim70~\mathrm{mT}$ with laser power of (d) 0.6~mW and (e) 5~mW.
The solid lines are the results of fitting using four equally spaced Lorentzians.
We estimate the area of each spectrum from the signal amplitude and line width obtained.
The vertical dashed lines indicate the resonance frequencies obtained by the fitting.
\label{fig3}
}\end{center}
\end{figure}

Next, we perform ODMR measurements to obtain $^{15}$N isotope effects on V$_\text{B}$ defects.
V$_\text{B}$ defects are created by helium ion implantation (acceleration voltage 30~keV, dose $1\times10^{15}~\mathrm{cm^{-2}}$) into flakes cleaved with Scotch tape.
The flakes are attached to silicon substrates (with a SiO$_2$ thickness of 90~nm).
We use the homemade confocal microscope~\cite{Misonou2020} with optimized optical filters for the photoluminescence (PL) of V$_\text{B}$ defects ($750\sim1000~\mathrm{nm}$).
A broadband microwave antenna with a copper wire soldered to a coplanar waveguide is used to mitigate unwanted distortions in the broad resonance spectrum of V$_\text{B}$ defects.
A magnetic field parallel to the optical ($z$) axis is applied by approaching a permanent magnet below the sample.

Figure~\ref{fig3}(a) shows the ODMR spectrum ($m_S=0\leftrightarrow-1$) of hB$^{14}$N at $B_z \sim 40~\mathrm{mT}$.
The broad signal consists of several closely overlapping Lorentzians (see SI).
The solid line is the fitted curve using Eq.~(\ref{odmrtot}) with $p_{15} = 0$.
It reproduces the experimental result well.
The parameters obtained by this fitting are $f_{-,0} = 2312\pm1~\mathrm{MHz}$, $C = 5.6\pm0.3\%$, $d\nu = 47\pm3~\mathrm{MHz}$, and $|^\text{(14N)}A_{zz}| = 43\pm1~\mathrm{MHz}$.
The obtained HFI parameter of $^{14}$N spin is consistent with previous works~\cite{Gottscholl2020,Murzakhanov2022,Gracheva2023,Gu2023} within a typical error of a few MHz.
Generally, it is impossible to determine the sign of the HFI parameter from this fitting.
From the positive zero-field splitting in the ground state~\cite{Gottscholl2020} and the spectral change at the ground state level anticrossing ~\cite{Shihao2023}, the sign of $^\text{(14N)}A_{zz}$ is most likely to be positive.
Note that $C$ and $d\nu$ depend on the measurement conditions, such as laser power and microwave amplitude~\cite{Drau2011}.

Next, we show the result of hB$^{15}$N in Fig.~\ref{fig3}(c).
The ODMR spectrum clearly consists of four dips.
Their separation is larger than in hB$^{14}$N.
These are the nitrogen isotope effects on V$_\text{B}$ defects.
The solid line is the fitted curve using Eq.~(\ref{odmrtot}) with $p_{15} = 1$ and reproduces the experimental result well.
The parameters obtained by this fitting are $f_{-,0} = 2308\pm1~\mathrm{MHz}$, $C = 11\pm0.3\%$, and $d\nu = 51\pm2~\mathrm{MHz}$, $|^\text{(15N)}A_{zz}| = 64\pm1~\mathrm{MHz}$.
As expected, the obtained magnitude of the HFI parameter of $^{15}$N spins, $|^\text{(15N)}A_{zz}|$, is $1.4$ times larger than the $|^\text{(14N)}A_{zz}|$ obtained above.
This observation is the central result of this work.

We describe the advantage of the obtained isotope effect, which sharpens the spectrum.
The maximum value of the slope of the ODMR spectrum increases with the amplitude and separation of each resonance.
Clearly, the maximum slope of the ODMR spectrum is larger for hB$^{15}$N than for hB$^{14}$N [Note that the vertical axis ranges are different in Fig.~\ref{fig3}(a) and (c)].
Since the magnetic field sensitivity is proportional to the slope, $^{15}$N isotope enrichment helps obtain high sensitivity.
For determining the nuclear spin isotope effects alone, the sensitivity gain is estimated to be about a factor of two when the analysis is performed, assuming that $C$ and $d\nu$, which depend on the measurement conditions~\cite{Drau2011}, are the same for both samples (see details in SI).
Furthermore, the reduced overlap of individual resonance lines directly implies improved fidelity in electron spin manipulation selective to nuclear spin states (see SI).
By applying operations with multi-frequency composite pulses~\cite{Gu2023}, it should be possible to excite most of the spectrum efficiently.

In addition, we measured hB$^{14+15}$N and obtained that the measured spectrum is consistent with the fitting using the HFI parameters and $p_{15} = 0.6$ [Fig.~\ref{fig3}(b)].
There are only slight undulations in the spectrum because it contains all signals from \#0 to \#3 [see Fig.~\ref{fig1}].
15N isotopic composition ratio of nearly 100\% is necessary to obtain isotope effects useful for quantum sensing.

Finally, we investigate the spectral changes induced by dynamic nuclear polarization due to the excited state level anticrossing ($B_z \sim 70$~mT)\cite{Jacques2009,Gao2022,Shihao2023}. 
In this situation, the angular momentum of the optically polarized electron spins in V$_\text{B}$ defects is transferred to the nuclear spins by flip-flops in the excited state~\cite{Gao2022,Shihao2023}.
Enhanced nuclear spin polarization can increase sensitivity by selectively increasing specific resonance signals~\cite{Drau2011,Sasaki2017}.
Figure~\ref{fig3}(d) is the ODMR spectrum of hB$^{15}$N at the magnetic field, where we observe the largest polarization.
Compared to Fig.~\ref{fig3}(c), there is clearly an increase in the signal on the high-frequency side and a decrease in the signal on the low-frequency side.
The polarization of $^{15}$N spins estimated from the area of spectra~\cite{Jacques2009,Gao2022,Shihao2023} is 16\%.
The precision of this estimation would be comparable to the signal amplitude and line width errors for each resonance (a few~\%).
Since it is enhanced to 27\% when the laser power is increased from 0.6~mW [Fig.~\ref{fig3}(d)] to 5~mW [Fig.~\ref{fig3}(e)], we conclude that this behavior is the result of the transfer of the polarization of the electron spin to the nuclear spins during optical transitions.
The trend of the observed change in resonance signals is opposite to that of conventional samples with the natural nitrogen composition ratio~\cite{Gao2022,Shihao2023}.
It indicates that the sign of the HFI parameter is opposite to $^\text{(14N)}A_{zz}$, i.e. $^\text{(15N)}A_{zz}=-64$~MHz, which is consistent with the different signs of the gyromagnetic ratio of $^{14}$N and $^{15}$N spin.
Even though such an apparent change appears, we also find polarization sign reversal at certain conditions that have not been reported before (see SI).
We leave the detailed polarization mechanism and the sign determination of $A_\text{zz}$ to future work.

In this work, we examine nitrogen isotope effects on V$_\text{B}$ defects in nitrogen isotopically enriched hBN.
We measure $^{15}$N isotopically enriched hBN crystals synthesized using the metathesis reaction under high pressure\cite{Chen2020,TaniguchiXXXX}.
In the hBN crystals with different $^{15}$N isotope composition, an isotope effect on phonon energy due to changes in the reduced mass are confirmed.
The magnitude of the HFI parameter of $^{15}$N spin is determined to be $64$~MHz from the fitting of ODMR spectra of V$_\text{B}$ defects created by helium ion implantation.
The demonstrated sharp spectrum of hB$^{15}$N is beneficial for achieving high sensitivity.
Further, when combined with $^{10}$B isotope enrichment techniques~\cite{Chen2020}, the sensitivity will be optimized by improving the coherence properties of V$_\text{B}$ defects\cite{Haykal2022}.
Sensor labeling with nitrogen isotopes may enable us to identify multiple sensor locations within a device stacked with two-dimensional materials.
The increased control fidelity and distinct optical polarization resulting from enhanced spectral separation would also make hB$^{15}$N useful as a polarization agent~\cite{Broadway2018,Jannin2019} and a platform for quantum information processing.
Furthermore, nitrogen isotope enrichment of hBN is essential in studying color centers other than V$_\text{B}$ defects, such as carbon-related defects\cite{Mendelson2020,Chejanovsky2021,Stern2023,Scholten2023}.
Our investigation, which reveals nitrogen isotope effects, is a vital step toward the design of hBN for quantum technologies.

We thank Kenji Watanabe (NIMS) for material preparation and Shu Nakaharai (TUT) for useful discussion, Kohei M. Itoh (Keio) for letting us use the confocal microscope system, and Ryota Akiyama (UTokyo) for supporting Raman measurement.
This work was partially supported by ``Advanced Research Infrastructure for Materials and Nanotechnology in Japan (ARIM)" (Proposal No. JPMXP1222UT1131) of the Ministry of Education, Culture, Sports, Science and Technology of Japan (MEXT), ``World Premier International Research Center Initiative on Materials Nanoarchitectonics (WPI-MANA)" supported by MEXT.
This work was supported by Grants-in-Aid for Scientific Research (KAKEN) Nos.~JP22K03524, JP19H00656, JP19H05826, JP23H01103, and JP23H02052, and Next Generation Artificial Intelligence Research Center at the University of Tokyo.

After the initial submission, we became aware of related works on V$_\text{B}$ defects in isotopically engineered hBNs~\cite{Gong2023,Clua2023}.

% % % Ref
% % % before submission --> *.bbl
% \bibliographystyle{apsrev4-1}
% \bibliography{main}% common bib file


\cleardoublepage
\onecolumngrid
% for Fig. S, Eq. (S)
\renewcommand{\thefigure}{S\arabic{figure}}
\renewcommand{\theequation}{S\arabic{equation}}
\renewcommand{\thetable}{S\arabic{table}}
\setcounter{figure}{0}
\setcounter{equation}{0}
\setcounter{table}{0}

\part{Supplemental Information of ``Nitrogen isotope effects on boron vacancy quantum sensors in hexagonal boron nitride''}

% ハミルトニアン説明
\section{Spin Hamiltonian}

In this section, we explain the spin Hamiltonian.
The spin Hamiltonian of the ground state of a V$_\text{B}$ defect would be given as,
\begin{align}
\hat{H} &= \hat{H}_\text{ZFS} + \hat{H}_\text{Ze} + \hat{H}_\text{Zn} + \hat{H}_\text{HFI} + \hat{H}_\text{QI}, \\
\hat{H}_\text{ZFS} &= D \hat{S}_z^2 + E_x (\hat{S}_y^2 - \hat{S}_x^2) + E_y (\hat{S}_x\hat{S}_y + \hat{S}_y\hat{S}_x), \\
\hat{H}_\text{Ze} &= \gamma_e \bm{B}_0 \cdot \hat{\bm{S}}, \\
\hat{H}_\text{Zn}  &= \sum_{j=1}^{3} -\gamma_{(j)} \bm{B}_0 \cdot \hat{\bm{I}}_{(j)}, \\
\hat{H}_\text{HFI} &= \sum_{j=1}^{3} \hat{\bm{S}} A_{\text{HFI},(j)} \hat{\bm{I}}_{(j)}, \\
\hat{H}_\text{QI}  &= \sum_{j=1}^{3} P_{p(j),(j)} \hat{I}_{p(j),(j)}^2 + P_{z,(j)} \hat{I}_{z,(j)}^2 + P_{o(j),(j)} \hat{I}_{o(j),(j)}^2,
\end{align}
where $z$ is the direction perpendicular to the hBN plane (the direction of the symmetry axis of the V$_\text{B}$ defect), $x$ and $y$ are the in-plane directions, $D$ is the zero-field splitting (ZFS) including the effects of electric field and strain, $\gamma_e = 28~\si{MHz/mT}$ is the gyromagnetic ratio of electron spin, $\bm{B}_0$ is the magnetic field vector, $E_x$ and $E_y$ are the strain parameters related to local electric field and crystal strain\cite{Dolde2011,Mittiga2018}, $j (=1,2,3)$ are labels of nearest-neighbor nitrogen sites, $\gamma_{(j)}$ is the gyromagnetic ratio of nitrogen nuclear spins, $A_{\text{HFI},(j)}$ is the hyperfine interaction (HFI) tensor, $\hat{I}_{k,(j)}$ is the nuclear spin operator in the $k$ direction, and $P_{k,(j)}$ is the nuclear quadrupole moment in the $k$ direction.

$\hat{H}_\text{ZFS}$ is the ZFS term, and $\hat{H}_\text{Ze}$ is the Zeeman term of the electron spin.
We assume that the strain terms take the same form as the NV center in diamond~\cite{Dolde2011,Mittiga2018}, which has the similar symmetry as the V$_\text{B}$ defect.
Typical parameter values for V$_\text{B}$ defects are $D\sim3450$~MHz and $E_x,E_y\sim 50$~MHz~\cite{Gottscholl2020,Gu2023,Ivdy2020,Gottscholl2021, Gao2022}.

$\hat{H}_\text{Zn}$ is the Zeeman term of nuclear spin, $\hat{H}_\text{HFI}$ is the HFI term, and $\hat{H}_\text{QI}$ is the nuclear quadrupole moment term.
They are based on the form of Ref.~\onlinecite{Gracheva2023}.
$p(j)$ is the direction from the vacancy (electron spin) to the nearest nitrogen site $j$, and the direction $o(j)$ is the cross-product direction of the $p(j)$ and $z$.
The gyromagnetic ratio is $\gamma_\text{14N} = 3.077~\si{kHz/mT}$ for $^{14}$N spin and $\gamma_\text{15N} = -4.316~\si{kHz/mT}$ for $^{15}$N spin.
The interactions with boron and nitrogen spins, except those of the nearest-neighbor nitrogen spins, are small and appear as a broadening of the ODMR line width~\cite{Haykal2022}, so we do not consider its details.

We introduce an approximation that is valid under quantum sensing conditions.
When a magnetic field is applied with sufficient strength in the direction of the symmetry axis ($\bm{B}_0 = B_z \bm{e}_z$), the effect of strain, which degrades the magnetic field sensitivity, can be ignored.
Specifically, this condition is given by $B_z \gg E_{x(y)}/\gamma_e$.
Except in the vicinity of the ground state level anticrossing ($D/\gamma_e \sim 125$~mT), the Hamiltonian can be approximated as,
\begin{align}
\hat{H}_\text{ZFS} &\approx D \hat{S}_z^2 \\
\hat{H}_\text{Ze} &= \gamma_e B_z \hat{S}_z, \\
\hat{H}_\text{HFI} &\approx \hat{S}_z \sum_{j=1}^{3} ( A_{zx,(j)} \hat{I}_{x,(j)} + A_{zy,(j)} \hat{I}_{y,(j)} + A_{zz,(j)} \hat{I}_{z,(j)} ),
\end{align}
where $A_{zx}$, $A_{zy}$, and $A_{zz}$ are the elements of the HFI tensor.
Within this approximation, the electron spin is quantized in the $z$ direction.

Then, we also introduce an approximation to the nuclear spin terms.
The HFI tensor consists of the dipole interaction and the Fermi contact interaction.
The element of the dipole interaction tensor between electron and nuclear spins is given by,
\begin{align}
^\text{dipole}A_{\alpha\beta} &= \frac{\mu_0}{4\pi} \frac{h \gamma_e \gamma_n} {\abs{\bm{r}}^3} [ 3 (\bm{e}_{\bm{r}}\cdot\bm{e}_{\alpha})(\bm{e}_{\bm{r}}\cdot\bm{e}_{\beta}) - \bm{e}_{\alpha}\cdot\bm{e}_{\beta}],
\end{align}
where $\alpha (= x,y,z)$ is the direction of the electron spin, $\beta (= x,y,z)$ is the direction of the nuclear spin, $h$ is the Plank constant, $\bm{r}$ is the position of the nuclear spin with respect to the electron spin, and $\bm{e}_j$ is a unit vector parallel to the $j$ direction.
Since the electron spin is quantized in the $z$ direction, only the $\alpha = z$ term needs to be considered.
Approximating the electron spin is localized at the vacancy position, $\bm{e}_{\bm{r}}\cdot\bm{e}_z=0$ is satisfied, and we obtain,
\begin{align}
^\text{dipole}A_{zz} &\approx -\frac{\mu_0}{4\pi} \frac{h \gamma_e \gamma_n} {\abs{\bm{r}}^3}, \\
^\text{dipole}A_{zx} &\approx 0, \\
^\text{dipole}A_{zy} &\approx 0.
\end{align}
The Fermi contact interaction $^\text{Fermi}A$ is a term arising from the overlapping of wave functions of electron and nuclear spins and is zero except for the isotropic component ($\alpha = \beta$).
Thus, the HFI term can be approximated as,
\begin{align}
\hat{H}_\text{HFI} &\approx \hat{S}_z \sum_{j=1}^{3} A_{zz,(j)} \hat{I}_{z,(j)}.
\end{align}
$\abs{A_{zz,(j)}}$ and typical line widths of the V$_\text{B}$ defects are around 40~MHz or larger.
Under typical experimental conditions, they are an order of magnitude larger than the nuclear spin's Zeeman effect and nuclear quadrupole moment.
Therefore, we neglect nuclear spin terms other than HFI and express the effective spin Hamiltonian as,
\begin{align}
\hat{H} &= D \hat{S}_z^2 + \gamma_e \bm{B}_z \hat{S}_z + \hat{S}_z \sum_{j=1}^{3} A_{zz,(j)} \hat{I}_{z,(j)}.
\end{align}
It corresponds to Eq.~(1) in the main text.
It is equivalent to ignoring the nuclear spin's Zeeman effect in Eq.~(8) of the Supplementary Information of Ref.~\onlinecite{Gao2022}.
In this condition, each nitrogen nuclear spin is quantized in the $z$ direction, and energy states according to their total quantum number $m_{I,tot}$ can be observed.

\section{Comparison of the ODMR spectra}

\begin{figure}
\begin{center}
\includegraphics{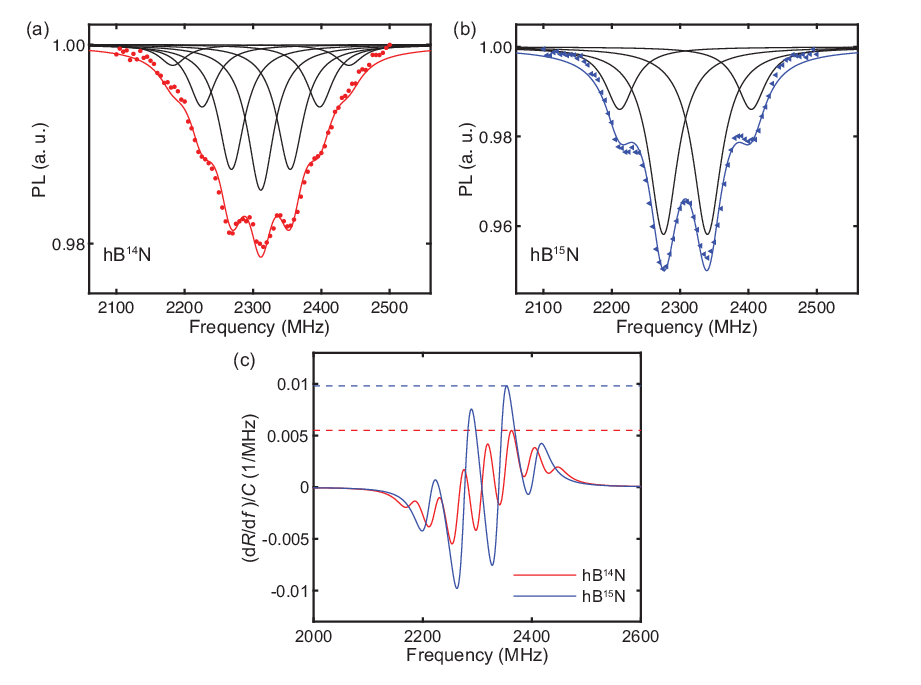}
\caption{
The enlarged images of (a) Fig.~3(a) and (b) Fig.~3(c) in the main text.
The solid lines show the fitting results.
(c) Comparison of ODMR spectral sharpness normalized by the signal amplitude $C$.
Both results were obtained using a $d\nu = 50$~MHz and experimentally obtained $A_\text{zz}$.
The horizontal dashed lines indicate the maximum slope (5.5~kHz$^{-1}$ for hB$^{14}$N and 9.8~kHz$^{-1}$ for hB$^{15}$N).
\label{figs1}
}\end{center}
\end{figure}

This section contains additional data related to Figs.~3(a) and (c) in the main text.
Figures~\ref{figs1}(a) and (b) are enlarged images of Figs.~3(a) and (c) in the main text, respectively.
Based on the fitting results, the signals of each resonance line are decomposed and shown.
The signal of hB$^{15}$N [Fig.~\ref{figs1}(b)] has a simpler spectrum with higher amplitude and narrower overall line widths than the conventional case [Fig.~\ref{figs1}(a)] reflecting that the number of included resonance lines is small and the separation of each is large.
Comparing the results of resolving each resonance (black lines) with the results of adding them together (blue/red line), we can see that the resonance overlap is smaller in hB$^{15}$N [Fig.~\ref{figs1}(b)] than in hB$^{14}$N [Fig.~\ref{figs1}(a)].
It directly results in the higher fidelity of nuclear spin state selective electron spin control in hB$^{15}$N.

The reasons for the slight deviation between the experimental spectra and the fitted result are not perfectly clear yet (also see following section).
It may be due to the polarization of nuclear spins or the frequency dependence of microwave power.

\section{Magnetic field sensitivity}

In magnetic field sensing using V$_\text{B}$ and NV centers, the shift in resonance frequency due to the Zeeman effect is determined from the ODMR spectral change. 
The magnetic field strength exerting a change in ODMR signal balanced by the photon counting shot noise is given by
\begin{align}
B_\text{z,min} &= \frac{1}{\sqrt{I_0 T}\abs{\frac{\partial R}{\partial B_z}}} 
\approx \frac{1}{\gamma_e\sqrt{I_0 T}\abs{\frac{\partial R}{\partial f_\text{esr}}} },
\end{align}
where $I_0$ is the photoluminescence (PL) intensity without a microwave, $T$ is the measurement duration, $R$ is the normalized ODMR spectrum, and $f_\text{esr}$ is the resonance frequency.
We assume that a sufficient bias magnetic field is applied in the direction of the symmetry axis and approximate the resonance frequency shift due to the magnetic field to correspond to the gyromagnetic ratio.
The magnetic field sensitivity is defined by $B_\text{z,min}$ per unit time, and is given by
\begin{align}
\eta_\text{B} &\approx \frac{1}{\gamma_e\sqrt{I_0}\abs{\frac{\partial R}{\partial f_\text{esr}}} } 
\propto \abs{\frac{\partial R}{\partial f_\text{esr}}}^{-1}.
\end{align}
The smaller $\eta_\text{B}$ is, the better the sensitivity becomes.
The sensitivity improves in proportion to the change in the ODMR spectrum with respect to the shift in the resonance frequency, i.e., the sharpness/slope of the spectrum.

When the ODMR spectrum appears as a single dip, the slope is determined by approximating it as a simple triangle as $\frac{\partial R}{\partial f_\text{esr}} \approx \frac{C}{d\nu}$, where $C$ and $d\nu$ are the signal amplitude and line width~\cite{Rondin2014}.

On the other hand, it is difficult to determine the slope by such a simple calculation when the several dips overlap, as in the case of the V$_\text{B}$ defect.
A straightforward method to obtain the slope is to differentiate the experimental or fitting results with respect to microwave frequency.
The slope obtained by differentiating the experimental fitting lines [Figs.~\ref{figs1}(a) and (b)] is three times greater for hB$^{15}$N than for hB$^{14}$N [Note that the vertical axis ranges are different in Fig.~\ref{figs1}(a) and (b)].
Figure~\ref{figs1}(c) shows the the spectra reproduced with $|A_\text{zz}|$ to compare the magnetic field sensitivity further.
To remove the influence of the different $C$ and $d\nu$, which depend on the microwave and laser strength, we set the line width as $d\nu = 50$~MHz and normalized the obtained slopes by $C$.
The maximum slope in hB$^{15}$N is about 1.8 times larger than in hB$^{14}$N.
It is a pure sensitivity gain caused by the nitrogen isotope effects.

\section{Nuclear spin polarization}

\begin{figure}
\begin{center}
\includegraphics{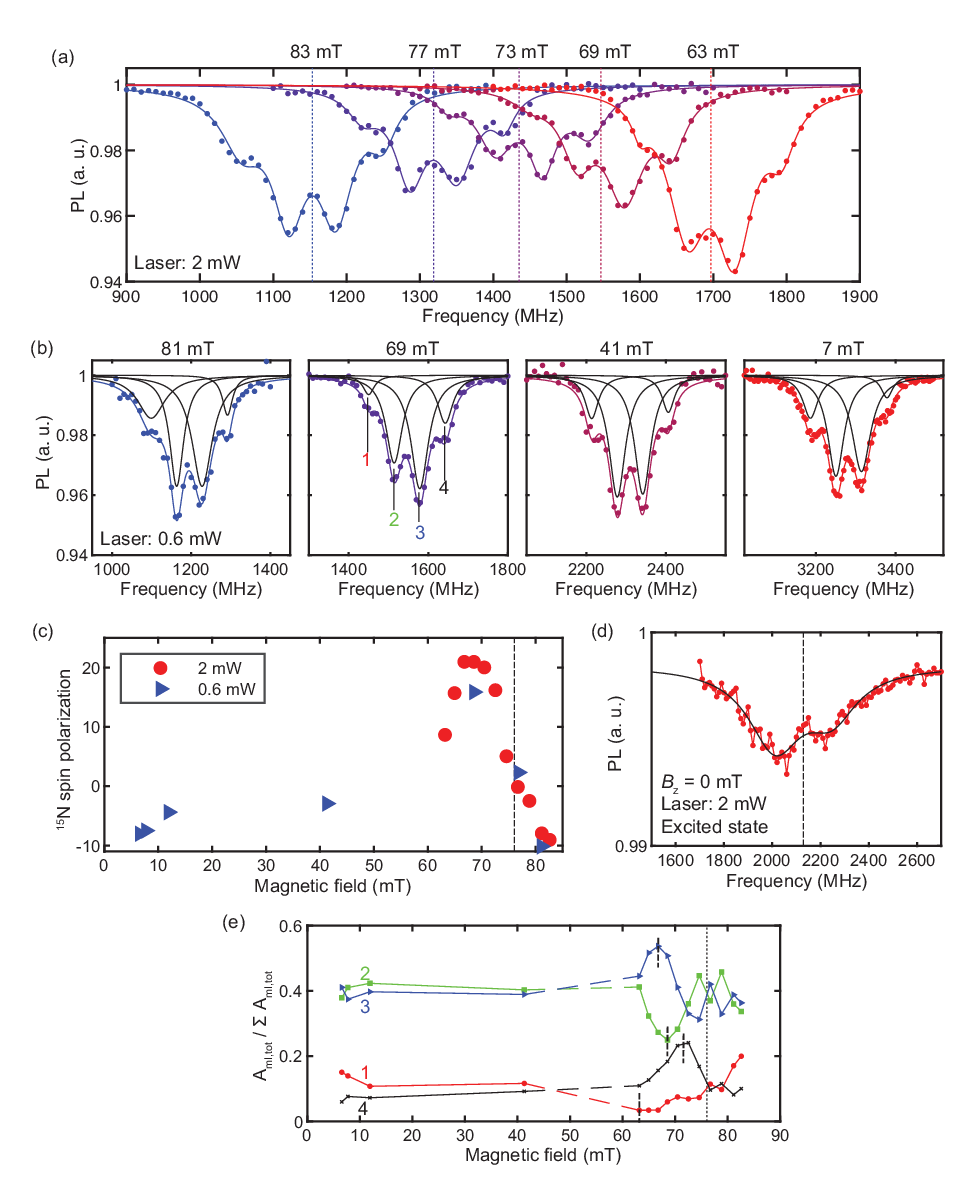}
\caption{
Magnetic field dependent nuclear spin polarization.
Magnetic field strength is estimated using $B_z = (D - f_{-,0})/\gamma_e$, where $D = 3466~\si{MHz}$ is the obtained zero-field splitting, and $f_{-,0}$ is the center frequency of the spectrum.
(a) ODMR spectra obtained at a laser power of 2~mW.
The vertical dotted line at the center of each spectrum and the number above it indicate the $f_{-,0}$ obtained by fitting and the corresponding magnetic field strength $B_z$, respectively.
(b) ODMR spectra obtained at a laser power of 0.6~mW.
(c) Estimated $^{15}$N spin polarization.
The vertical dashed line indicates the condition of excited state level anticrossing (ESLAC).
(d) ODMR spectrum of the excited state at a zero magnetic field.
(e) Estimated area of each resonance normalized by the total area.
The data below 50~mT are only those obtained with a laser power of 0.6~mW, and the data above 50~mT are only those acquired with a laser power of 2~mW. 
For visibility, the data points are connected by straight or dashed lines.
Each resonance is labeled 1, 2, 3, and 4, starting from the lowest frequency.
\label{figs2}
}\end{center}
\end{figure}

Here we show the additional data of Figs.~3(d) and (e) in the main text.

In the excited state leve anticrossing (ESLAC) condition, the angular momentum of the optically polarized electron spins in V$_\text{B}$ defects is transferred to the nuclear spins by flip-flops.
The direction of nuclear spin polarization depends on the sign of the zero-field splitting of the excited state.
The sign of the zero-field splitting in the excited state is likely to be positive as in the ground state since the resonance signal corresponding to the same nuclear spin state is enhanced at the level anticrossing of the ground and excited states in a previous study~\cite{Shihao2023}.
Considering that the anticrossing levels in the condition are $m_S = 0$ and $-1$ states and electron spin is optically polarized to $m_S = 0$ state~\cite{Gottscholl2020}, the nuclear spin polarization is positively increased by the flip-flop.

Figure~\ref{figs2}(a) shows the obtained spectra at a laser power of 2~mW at magnetic fields of 83~mT, 77~mT, 73~mT, 69~mT, and 63~mT from left to right.
As shown in the figure, the ODMR spectrum consists of four resonance lines. 
Each is named resonances 1, 2, 3, and 4, in descending order of frequency [see the graph for 69~mT in Fig.~\ref{figs2}(b)].
We observe a property that biases the spectrum toward the high-frequency side around 70~mT.
This is the opposite behavior of conventional non isotope-controlled hBN.
Figure~\ref{figs2}(c) shows the $^{15}$N spin polarization estimated by~\cite{Gao2022,Shihao2023},
\begin{align}
\text{Polarization} = \frac{ \sum m_{I,tot} A_{m_{I,tot}} }{ \frac{3}{2}\sum{ A_{m_{I,tot}} } },
\label{pol}
\end{align}
where $A_{m_{I,tot}}$ is the area of the spectrum belonging to the $m_{I,tot}$ state, estimated from the product of signal amplitude and line width obtained by fitting each spectrum.
The summation symbols in the denominator and numerator are for the possible $m_{I,tot}$ states.
We analyzed resonances 1, 2, 3, and 4 as corresponding to $m_{I,tot} = -3/2$, $-1/2$, $1/2$, and $3/2$, respectively.
The polarization reaches a maximum of around 70~mT.
This condition is close to the ESLAC estimated to be 76~mT from the zero-field splitting of $2130~\si{MHz}$ obtained from the ODMR spectrum of the excited state measured at zero field [Fig.~\ref{figs2}(d)].

The polarization is enhanced at lower field conditions than estimated from the excited state. 
Figure~\ref{figs2}(e) shows the estimated area of resonance 1, 2, 3, and 4.
The maximum and minimum of the area of each resonance occur at different magnetic field conditions (see short vertical dashed lines).
This behavior may be due to the fact that the anticrossing condition corresponding to the nuclear spin state shifts according to the $A_\text{zz}$ in the excited state.
The observed condition shift to lower field can occur when the zero-field splitting and $A_\text{zz}$ in the excited state have the same sign. 
Since the ground and excited states will have different spin wave function distributions in real space, the respective $A_\text{zz}$ can have different signs.

We show the results of ODMR spectra obtained under broader magnetic field conditions, the polarization estimated from them, and the areas of each resonance in Figs.~\ref{figs2}(b), (c), and (e), respectively.
As with the case where the laser power is 2~mW, polarization increased around 70~mT.

In contrast, interestingly, the negative polarization is obtained at low fields and conditions beyond ESLAC.
The minimum polarization is about $-10$\% and its magnitude is comparable to the polarization around 70~mT.
These estimates are due to the large area ratio of the lowest frequency resonance 1 and the small area ratio of the highest frequency resonance 4, except around 70~mT.
Although the resonance 3 and 4, which belong to the same polarization direction, increase and decrease with similar field strength, the resonance 1 and 2 do not appear to be correlated except at about 70~mT [see Fig.~\ref{figs2}(e)].
Thus, this behavior outside of the 70 mT region can't be explained by nitrogen nuclear spin polarization alone.

\begin{figure}
\begin{center}
\includegraphics{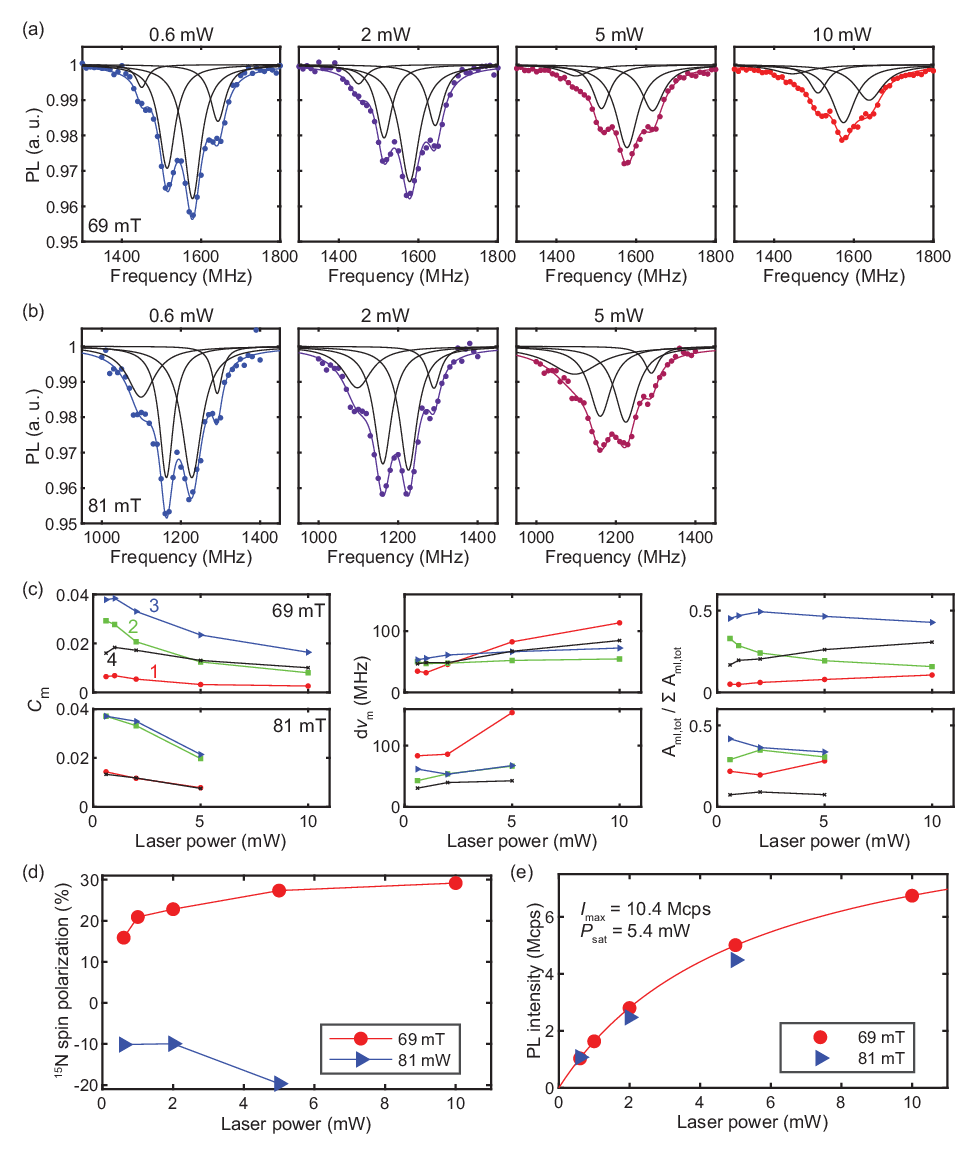}
\caption{
Optical power dependent nuclear spin polarization.
(a) ODMR spectra obtained at a magnetic field of 69~mW.
(b) ODMR spectra obtained at a magnetic field of 81~mW.
(c) Obtained fitting parameters and estimated area of each resonance.
$C_m$ and $d\nu_m$ are the signal amplitude and line width of the resonance $m$ [see Fig.~\ref{figs2}(b)].
(d) Estimated $^{15}$N spin polarization.
(e) Saturation of the PL intensity.
The solid line is the fitted result with $I_\text{max} \frac{P}{P+P_\text{sat}}$, where $P$ is the laser power.
\label{figs3}
}\end{center}
\end{figure}

We also investigated the spectral change with laser power.
Figures~\ref{figs3}(a) and (b) are spectra obtained at a magnetic field of 69~mT and 81~mT at several different laser powers, respectively.
The obtained fitting parameters and estimated polarization are shown in Fig.~\ref{figs3}(c) and (d), respectively.
At 69~mT, the polarization increases smoothly.
It is a typical behavior of optical nuclear spin polarization at the ESLAC~\cite{Gao2022,Shihao2023}.
In addition, polarization saturation appears to occur at weaker laser power than PL intensity saturation [Fig.~\ref{figs3}(e)].

On the other hand, at 81~mT, the polarization is estimated to be negative.
It doubles at 5~mW, although no significant change appears at 0.6~mW and 2~mW.
More interestingly, this change in polarization is attributed to a dramatic increase in the line width of the lowest frequency resonance 1 [see Fig.~\ref{figs3}(c)].
Although a similar characteristic is observed at 69~mT, the polarization is not negative due to an amplitude reduction, consistent with ESLAC, and gradual change of the line width of resonance 1.

Here we organize the above results.
First, it was confirmed that the ODMR spectrum of hB$^{15}$N is biased in a different frequency direction from that of hB$^{14}$N under conditions close to ESLAC.
This nuclear spin polarization depends on the laser power, which suggests that it is related to the optical polarization or excited state of the electron spins.
From these results, the polarization increase is most likely due to ESLAC, and the sign of the ground state $A_\text{zz}$ is expected to be opposite to that of hB$^{14}$N.
Second, on the other hand, it is observed that under conditions outside the ESLAC, the estimated polarization direction is inverted.
This phenomenon has never been observed for conventional hBN.
This trend appears because the line width is broadened while the amplitude of the lowest frequency resonance signal is maintained.
The ESLAC-mediated spin polarization mechanism cannot explain this behavior.
Since there are some unexplained aspects of polarization behavior, we will pursue the causes of these phenomena, including the determination of the sign of $A_\text{zz}$, as a topic for future research.

% % Ref
% \bibliographystyle{apsrev4-1}
% \bibliography{main}% common bib file
% % before submission --> *.bbl


\end{document}